\documentclass[12pt]{article}
\usepackage{amsfonts}
\usepackage{bm}
\usepackage{amsmath}
\usepackage{epsfig}
\usepackage[mathscr]{euscript}

\usepackage{slashed}

\newcommand{\bmat}{\left(\begin{array}}
\newcommand{\emat}{\end{array}\right)}

\def\gtrsim{\mathrel{\raise.3ex\hbox{$>$\kern-.75em\lower1ex\hbox{$\sim$}}}}

\def\a{\alpha}
\def\ap{\alpha^{\prime}}

\def\-{\hphantom{-}}
\def\ov{\overline}
\def\un{\underline}
\def\s2{\frac{1}{\sqrt2}}

\def\mg{m_{3/2}}
\def\mg2{m^2_{3/2}}

\def\Dsl{\,\raise.15ex\hbox{/}\mkern-13.5mu D} 

\def\be{\begin{equation}}
\def\ee{\end{equation}}
\def\bea{\begin{eqnarray}}
\def\eea{\end{eqnarray}}

\newcommand{\nn}{\nonumber}

\begin{document}

\pagestyle{plain}

\makeatletter
\@addtoreset{equation}{section}
\makeatother
\renewcommand{\theequation}{\thesection.\arabic{equation}}
\pagestyle{empty}
\begin{center}
\ \

\vskip .5cm
\LARGE{\LARGE\bf ${\cal N}=1$ Supersymmetric Double Field Theory and the generalized Kerr-Schild Ansatz
 \\[10mm]}
\vskip 0.3cm

\large{Eric Lescano$^\dag$ and Jes\'us A. Rodr\' iguez$^*$
 \\[6mm]}

{\small  $^\dag$ Instituto de Astronom\'ia y F\'isica del Espacio (IAFE-CONICET-UBA)\\ [.01 cm]}
{\small\it Ciudad Universitaria, Pabell\'on IAFE, 1428 Buenos Aires, Argentina\\ [.3 cm]}
{\small  $^*$ Departamento de F\'isica, FCEyN, Universidad de Buenos Aires (UBA) \\ [.01 cm]}
{\small\it Ciudad Universitaria, Pabell\'on 1, 1428 Buenos Aires, Argentina\\ [.5 cm]}

{\small \verb"elescano@iafe.uba.ar, jarodriguez@df.uba.ar"}\\[1cm]

\small{\bf Abstract} \\[0.5cm]\end{center}
 
We construct the ${\cal N} = 1$ supersymmetric extension of the generalized Kerr-Schild ansatz in the flux formulation of Double Field Theory. We show that this ansatz  is compatible with ${\cal N}=1$ supersymmetry as long as it is not written in terms of generalized null vectors. Supersymmetric consistency is obtained through a set of conditions that imply linearity of the generalized gravitino perturbation and unrestricted perturbations of the generalized background dilaton and dilatino. As a final step we parametrize the previous theory in terms of the field content of the low energy effective $10$-dimensional heterotic supergravity and we find that the perturbation of the $10$-dimensional vielbein, Kalb-Ramond field, gauge field, gravitino and gaugino can be written in terms of vectors, as expected.

\newpage
\setcounter{page}{1}
\pagestyle{plain}
\renewcommand{\thefootnote}{\arabic{footnote}}
\setcounter{footnote}{0}

\tableofcontents
\newpage

\section{Introduction}

General Relativity is a very non-linear theory and many efforts were made in order to find exact solutions. The rotating black hole solution (the Kerr black hole) \cite{Kerr} was the initial construction of a very simple and powerful ansatz called the Kerr-Schild ansatz \cite{KS}. This ansatz consists in an exact and linear perturbation of a background metric tensor $g_{o \mu \nu}$ of the form,
\bea
g_{\mu \nu} = g_{o \mu \nu} + \kappa l_{\mu} l_{\nu} \, ,
\label{sugrametric}
\eea
such that $\kappa$ is an arbitrary parameter that allows to quantify the order of the perturbation and $l_{\mu}$ is a null vector with respect to $g_{\mu\nu}$ and $g_{o\mu\nu}$ \textit{i.e.}
\bea
\label{null}
g^{\mu \nu} l_{\mu} l_{\nu} & = & g_{o}^{\mu \nu} l_{\mu} l_{\nu} = 0 .
\eea
With this assumption, the exact inverse to (\ref{sugrametric}) is
\bea
g^{\mu \nu} = g_{o}^{\mu \nu} - \kappa l^{\mu} l^{\nu} \, .
\eea
If we ask for linearity in the e.o.m of $g_{\mu \nu}$ \cite{Gurses}, then $l_{\mu}$ is also a geodesic vector with respect to the background metric
\be
g^{\mu \nu} l_{\mu} \nabla_{o \nu} l_{\rho} = 0  \, ,
\label{geodesic}
\ee
where $\nabla_{o}$ is a compatible and torsion-free covariant derivative using the Levi-Civita connection that depends on $g_{o \mu \nu}$.
 
The Kerr-Schild formalism has been successful in different contexts of theoretical physics. It can be used to describe not only the Kerr black hole but also the Myers and Perry black hole \cite{Perry}, Einstein-Gauss-Bonnet gravity \cite{Anabalon}, Einstein-Lovelock gravity \cite{Kastor}, a perturbative duality between gauge and gravity theories referred as Classical Double Copy \cite{Dcopy} and it has recently been applied \cite{KL} in the context of Double Field Theory (DFT). In this work DFT \cite{Siegel,DFT,Park_DFT,Heterotic_DFT,Rev} is understood as a rewriting of a classical $d$-dimensional supergravity in a more general way such that the generalized version of the supergravity is manifestly invariant under the action of $G=O(d,d)$. As $G$ is closely related to a symmetry of String Theory, DFT is often applied to reformulate supergravities whose bosonic field content includes a $2$-form $b_{\mu \nu}$ (or Kalb-Ramond field) and a scalar field $\phi$ (or dilaton) in addition to the metric tensor. These fields conform the universal NS-NS sector of all the formulations of String Theory. The generalized field content of DFT can accommodate the supergravity field content in multiplets of the duality group and a generalized notion of geometry can be defined.    

One of the most distinctive features of DFT is that the space-time coordinates of the $d$-dimensional supergravity must be doubled,
\be
X^{M}=(x^{\mu},\tilde x_{\mu}) \, ,
\ee
where $M=0,\dots,2d-1$ and $X^{M}$ is a generalized coordinate in the fundamental representation of $G$. The addition of the coordinates $\tilde x_{\mu}$ forces the appearance of the strong constraint,
\be
\partial_{M} \star \partial^{M} \star = 0\qquad \, , \qquad \partial^{M}\partial_{M}\star = 0
\label{SC}
\ee
where $\star$ means any combination of fields or parameters of the theory and the contractions are done with the $G$-invariant metric $\eta_{MN}$. From a stringy point of view, the constraint (\ref{SC}) is related to the Fourier transformation of the Level Matching Condition when winding modes are admitted, and written in a duality covariant way \cite{Mariana}. The dynamical background metric of DFT is the generalized metric $H_{o MN}$, which is a multiplet and an element of $G$, \textit{i.e.}
\bea
H_{o M P} \eta^{P Q} H_{o N Q} = \eta_{M N} \, ,
\eea
parametrized by the background metric tensor $g_{o \mu \nu}$ and the background Kalb-Ramond field $b_{o \mu \nu}$.

The generalized Kerr-Schild ansatz was defined by K. Lee in \cite{KL} as an exact and linear perturbation of the generalized background metric with the following form
\bea
H_{MN} = H_{o MN} + \kappa \bar{K}_{M} K_{N} + \kappa {K}_{M} \bar{K}_{N}  \, ,
\label{DFTKS}
\eea
where $\bar{K}_{M}= \bar{P}_{M}{}^{N} \bar{K}_{N}$ and $K_{M}= {P}_{M}{}^{N} {K}_{N}$ are a pair of generalized null vectors 
\bea
\eta^{MN} \bar{K}_{M} \bar{K}_{N} & = & \eta^{MN} K_{M} K_{N} =  \eta^{MN} \bar{K}_{M} K_{N} = 0 \,
\label{nulldft}
\eea 
that satisfy
\bea
{\bar K}^{P} \nabla_{o P} K^{M} + K_{P} \nabla_{o}^{M}{\bar K}^{P} - K^{P} \nabla_{o P}{\bar K}^{M} & = & 0 \, , \nn \\ {K}^{P} \nabla_{o P} {\bar K}^{M} + {\bar K}_{P} \nabla_{o}^{M}{K}^{P} - {\bar K}^{P} \nabla_{o P}{K}^{M} & = & 0 \, , 
\label{geodesicdft}
\eea
where $\bar{P}_{MN}=\frac12(\eta_{MN} + H_{MN})$ and ${P}_{MN}=\frac12(\eta_{MN} - H_{MN})$ are used to project the $O(d,d)$ indices and $\nabla_{o M}$ is a generalized covariant derivative. Relying on the previous conditions, the e.o.m of the generalized metric can be linearized in a similar fashion to (\ref{geodesic}). The ansatz (\ref{DFTKS}) and the conditions (\ref{nulldft}) and (\ref{geodesicdft}) were proposed and analyzed in the semi-covariant formalism of DFT and a perturbation for the generalized dilaton $d$ (parameterized by the 10-dimensional dilaton $\phi$) was also considered. 

\subsection{Main Results}

The main goal of this work is to construct the ${\cal N}=1$ supersymmetric extension of the ansatz (\ref{DFTKS}) in the flux formalism of DFT \cite{Flux,SDFT}. As we include generalized fermionic degrees of freedom, we are forced to work in the generalized background frame formalism and fix the space-time dimension. Particularly we consider $d=10$. Since we are dealing with the same degrees of freedom as the supergravity limit of Heterotic String Theory, we let the inclusion of gauge fields in our setup and the starting point is a ${\cal N}=1$ DFT with $G=O(10,10+n)$ invariance, where $n=496$ is the dimension of the heterotic gauge group \cite{Heterotic_DFT}. We consider the leading order terms in fermions and show that ${\cal N}=1$ supersymmetry is compatible with the generalized Kerr-Schild ansatz as long as it is not written in terms of generalized null vectors. 

The most general linear perturbation of the generalized frame is,
\bea
E_{M}{}^{\ov A} = E_{oM}{}^{\ov A} + \frac12 \kappa E_{o M}{}^{\underline B} {\Delta}_{\underline B}{}^{\ov A}\, \nn \\ E_{M}{}^{\underline A} = E_{oM}{}^{\underline A} - \frac12 \kappa E_{o M}{}^{\ov B} {\Delta}^{\underline A}{}_{\ov B}{}
\label{pert1}
\eea
where $A=(\underline A,\overline A)$ are indices in $O(9,1)_L \times O(1,9+n)_R$ respectively and $\Delta_{A}{}^{B}$ is a mixed-projected perturbation that satisfies,
\bea
\Delta_{\ov A \ov{B}} = \Delta_{\underline A \underline{B}} & = & 0\, , \nn \\
\Delta \eta^{-1} \Delta & = & 0 \, , 
\label{square}
\eea
in order to be consistent with the constraints of DFT. We find that (\ref{pert1}) cannot be written in terms of generalized null vectors when supersymmetry is considered and therefore conditions (\ref{geodesicdft}) are not available to simplify the perturbation of the generalized Ricci scalar and/or the e.o.m of the generalized frame. We perturb the generalized background dilaton, gravitino and dilatino in the following way,
\bea
d & = & d_{o} + \kappa f\, , \qquad f = \sum _{n}\kappa^{n}f_{n} \\ \Psi_{\overline A} & = & \Psi_{o \overline A} + \kappa \Theta_{\overline A}\, , \quad \Theta_{\overline A} = \sum _{n}\kappa^{n}\Theta_{n \overline A}  \\
\rho & = & \rho_{o} + \kappa g\, , \qquad g = \sum _{n}\kappa^{n}g_{n} \, ,
\label{pert2}
\eea
where $n\geq 0$. With the previous setup we find that ${\cal N}=1$ supersymmetry only restricts the generalized gravitino expansion,
\bea
\Theta_{n} & = & 0 \quad , \quad n>1 \, ,
\label{gravitinoc}
\eea
while the perturbations of the generalized dilatino and dilaton remain unrestricted. Condition (\ref{gravitinoc}) forces the following supersymmetric consistency conditions,
\bea
\delta\cdots\delta\left(\Theta_1\right) = \delta\cdots\delta\left(\Theta_2\right) = 0 \, ,
\label{conpsi}
\eea
with $\delta$ a generic symmetry transformation. 

As a final step we parametrize the generalized perturbations in terms of the heterotic supergravity field content and we find,
\bea
\label{paramsugra}
g_{\mu \nu} & = & g_{o \mu \nu} + \frac{\kappa}{1+\frac12 \kappa l.\bar{l}} l_{(\mu} \bar{l}_{\nu)} \nn \\ b_{\mu \nu} & = & b_{o \mu \nu} - \frac{\kappa}{1+\frac12 \kappa l.\bar{l}} l_{[\mu} (\bar{l}_{\nu]} - \frac{1}{\sqrt{2}} j_{i} A_{\nu]}{}^{i} ) \nn \\ \phi & = & \phi_{o} + \kappa f \nn \\
\psi_{a} & = & \psi_{oa} - \frac{\kappa}{2+\kappa l.\bar{l}} \bar{l}_{(a} l_{b)} \psi_{o}^{b} \nn \, , \\ \lambda & = & \lambda_{o} + \frac{\kappa}{2} g \, ,
\eea
where $e_{\mu a}$ is a $10$-dimensional vielbein, $\bar{l}_{a}=e^{\mu}{}_{a} \bar{l}_{\mu}$ and ${l}_{a}=e^{\mu}{}_{a} {l}_{\mu}$ are a pair of vectors and $\psi_{a}$ and $\lambda$ are the 10-dimensional gravitino and dilatino of the effective heterotic supergravity. The indices $\mu=0 \dots 9$ and $a=0 \dots 9$ are space-time and $O(1,9)$ Lorentz indices respectively. In (\ref{paramsugra}) $n$ is not fixed by supersymmetry as happens in DFT. The ordinary Kerr-Schild ansatz is recovered when $l_{a}=\bar{l}_{a}$. The remaining fields of the effective heterotic supergravity are 
\bea
 A_{\mu i} & = & A_{o \mu i} + \frac{1}{\sqrt{2}} \frac{\kappa}{1+\frac12 \kappa l.\bar{l}} l_{\mu} j_{i} \nn \\ \chi_{i} & = & \chi_{o i} - \frac{\kappa}{2} l_{b} j_{i} \psi^{b}_{o} \, , 
\eea
where $A_{\mu i}$ is a $10$-dimensional gauge connection, $\chi$ is a $10$- dimensional gaugino and $j_{i}$ parameterizes the perturbations when gauge fields are included.  For the parametrization of the generalized perturbations we consider
\bea
\label{dec 0}
\Delta_{\underline A \overline B} = (\frac{2}{1+\frac12 \kappa l.\bar{l}} l_{(a} \bar l_{b)} \delta^{ab}_{\underline a \overline b}, \Delta_{ai} \delta^{ai}_{\underline a \overline i}) \\ 
\Theta_{0\overline A} =  (-\frac{1}{2+\kappa l.\bar{l}} l_{(a} \bar{l}_{b)} \psi^{b}_{0} \delta_{\overline a}^{a},\Theta_{0i} \delta^{i}_{\bar i}) \, ,
\eea
where $\Delta_{ai}=l_{a} j_{i}$ and
${\cal N}=1$ supersymmetric consistency forces  $\Theta_{0i}= - \frac12 l_{b} j_{i} \psi_{o}^{b}$.

A very interesting aspect of (\ref{dec 0}) is that the supersymmetric extension of the generalized Kerr-Schild formalism can be parametrized in terms of a pair of vectors. The supersymmetric consistency constraints of DFT (\ref{conpsi}) can be understood as some extra conditions on the expansion of $\psi_{a}$. The transformation rule of $l_{a}$ and $\bar l_{a}$ is
\bea
\delta l_{a} & = & \xi^{\mu} \partial_{\mu} l_{a} + l_{b} \Lambda^{b}{}_{a} \, , \\
\delta \bar l_{a} & = & \xi^{\mu} \partial_{\mu} \bar l_{a} + \bar l_{b} \Lambda^{b}{}_{a}
\eea
where $\Lambda_{ab}$ parametrizes a $O(1,9)$ Lorentz symmetry and $\xi_{\mu}$ parametrizes $10$-dimensional diffeomorphisms. The previous conditions are stronger than the usual geodesic equation, but in this case the e.o.m of $g_{\mu \nu}$ is no more linear in $\kappa$ due to the $10$-dimensional dilatonic and fermionic perturbations.   

This work is organized as follows: In section \ref{DFT} we introduce the field content, the symmetries and the action principle of ${\cal N}=1$ DFT for background fields. Section \ref{GKS} is dedicated to explore the supersymmetric extension of the generalized Kerr-Schild ansatz. First we include finite perturbations on the background field content. Then we discuss the supersymmetric consistency conditions and write schematically the action principle and the equations of motion. In section \ref{Param} we parametrize the theory in terms of the field content of the $10$-dimensional heterotic supergravity and find the extra supersymmetric conditions that are necessary for consistency. We discuss about the kind of solutions that can be found with the present formalism in section \ref{app}. As an explicit example we analyze the $d=10$ gaugino condensation in the fundamental charged heterotic string. Finally, in section \ref{Conclu} we present the conclusions of the work and some future directions to explore.   

\section{${\cal N}=1$ Supersymmetric Double Field Theory}
\label{DFT}

${\cal N}=1$ supersymmetric DFT is defined on a double space with coordinates $X^{M}$ which transforms under the fundamental representation of the symmetry group $G=O(10, 10+n)$, with $M=0, \dots, 19+n$, and $n$ the dimension of the gauge group. For instance $n=496$ if we want to encode a low energy description of heterotic supergravity in a T-duality covariant framework. The theory is invariant under a global $G$ symmetry which infinitesimally reads
\bea
\delta_G V_M = V_{N} h^{N}{}_{M} \, ,
\label{duality}
\eea
where $V_M$ is a generic $G$-multiplet and $h \in O(10, 10+n)$ is the $G$-parameter. The invariant metric of $G$ is $\eta_{MN} \in G $ and $G$-invariance imposes
\bea
h_{MN} = - h_{NM} \, ,
\eea
where we use $\eta$ and $\eta^{-1}$ in order to lower and raise all the G-indices.

Another symmetry of the theory are generalized diffeomorphisms, generated infinitesimally  by $\xi^{M}$  through the generalized Lie derivative, defined by 
\bea
\hat{\cal L}_\xi V_M = \xi^{N} \partial_N V_M + (\partial_M \xi^N - \partial^N \xi_{M}) V_N + f_{MNP} \xi^{N} V^{P} + t \partial_{M} \xi^{M} \, ,
\eea 
where $V_M$ is an arbitrary generalized tensor, $t$ is a weight constant and $f_{MNP}$ are the generalized version of the structure constants that satisfy
\bea
f_{{ MNP}}=f_{[{MNP}]}\, , \qquad f_{[ MN}{}^{ R}f_{{P}] R}{}^{ Q}=0\, . \label{consf}
\eea

The theory is also invariant under a local double Lorentz  ${\cal H}=O(9,1)_L\times O(1, 9+n)_R$ symmetry generated infinitesimally by a generalized parameter $\Gamma_{AB}$ where ${A}=(\underline{A},\overline A)$  splitting into $O(9,1)_L$  and $O(1,9+n)_R$ vector indices, $\underline{A}={\un a}=0,\dots , 9$  and   $\overline A=(\ov a, \ov i)=0,\dots , 9+n$, {\textit i.e.}, 
\bea
\delta_{\cal H} V_A = V_B \Gamma^{B}{}_{A} \, ,
\label{Lambda}
\eea
for a generic ${\cal H}$-vector. The $\cal H$-invariance of $\eta_{AB}$ imposes $\Gamma_{AB} = - \Gamma_{BA} \,$.

Supersymmetry is parameterized by an infinitesimal generalized Majorana fermion $\epsilon$ which behaves as a spinor of $O(9,1)_L$. We work at leading order in fermions, such that supersymmetric transformation of bosons are at most quadratic in fermions, and supersymmetric transformation of fermions are linear in fermions. The explicit transformation rules will be discussed later.

The fundamental background fields of the theory consist in a generalized frame $E_{oM}{}^{A}$ parameterizing the coset $\frac{G}{\cal H}=\frac{O(10,10+n)}{O(9,1)_L\times O(1,9+n)_R}$, and a generalized dilaton field $d_{o}$. The action of the symmetry groups on these fields is

\begin{center}
\begin{tabular}{||c c c c c||}
\hline
& $G$ & ${\cal H}_{L}$ & ${\cal H}_{R}$ & Diff \\ [0.5ex]
\hline\hline
$E_{oM}{}^{A}$ & $G$-vector & ${\cal H}_{L}$-vector & ${\cal H}_{R}$-vector & tensor \\
\hline
$d_{o}$ & $G$-invariant & ${\cal H}_{L}$-invariant & ${\cal H}_{R}$-invariant & scalar($t=-\frac12$)\, \\ [1ex]
\hline
\end{tabular}
\end{center}

Consistency of the construction requires 
constraints which restrict the coordinate dependence of fields and gauge parameters. The strong constraint
\be
\partial_{ M} \partial^{ M} \star = 0 \ , \ \ \ \ \ \partial_{ M} \star \ \partial^{ M} \star = 0 \, , \ \ \ \ \
f_{ {MN}}{}^{ P}\partial_{ P}\star =0\, ,\label{StrongConstraint}
\ee
where $\star$ refers to products of fields, will be assumed throughout. This constraint  locally removes   the field dependence  on $10+n$ coordinates, so that fermions can be effectively defined in a $10$-dimensional tangent space. 

The frame-formulation of DFT demands the existence of two constant, symmetric and invertible ${\cal H}$-invariant metrics $\eta_{{AB}}$ and ${H}_{{AB}}$. The former is used to raise and lower the indices that are rotated by ${\cal H}$ and the latter is constrained to satisfy 
\bea
{H}_{A}{}^{C}{H}_{C}{}^{B} = \delta_{A}^{B}\, .
\eea 
The generalized background frame $E_{o}^{M}{}_{A}$ is constrained to relate the metrics $\eta_{{AB}}$ and $\eta_{ {MN}}$ and defines a generalized background metric ${ H}_{oMN}$ from ${H}_{AB}$
\be
\eta_{AB} = E_{o}^{M}{}_{A}\eta_{MN}E_{o}^{N}{}_{B}\, , \quad {H}_{oMN} = E_{oM}{}^{A}{H}_{AB}E_{oN}{}^{B} \, .
\ee
${H}_{oMN}$ is also an element of $O(10,10+n)$, \textit{i.e.} 
\be
{H}_{oMP}\eta^{PQ}{H}_{oQN} = \eta_{MN} \, . \label{GMconstraint}
\ee
It is convenient to introduce the projectors
\bea
P_{oMN} = \frac{1}{2}\left(\eta_{MN} - {H}_{oMN}\right)  \ \ {\rm and} \ \
\ov{P}_{oMN} = \frac{1}{2}\left(\eta_{MN} + {H}_{oMN}\right)\ ,
\eea
which satisfy the usual properties 
\bea
&{\overline{P}_o}_{{M Q}} {\overline{P}_o}^{ Q}{}_{ N}={\overline{P}_o}_{{M N}}\, , &\quad {P_o}_{{M Q}} {P_o}^{Q}{}_{ N}={P_o}_{{M N}}, \nn\\
&{P_o}_{{M  Q}}{\overline{P}_o}^{Q}{}_{ N} = {\overline{P}_o}_{ {M Q}}  {P_o}^{ Q}{}_{ N} = 0\, ,  &\quad {\overline{P}_o}_{{MN}} + {P_o}_{{M N}} = \eta_{{M N}}\,,
\eea
and the same can be done with $\eta_{AB}$ and $H_{AB}$ to define ${P}_{oAB}$ , $\ov{P}_{o {AB}}$. We use the convention that ${P}_{o {AB}}$ , $\ov{P}_{o{AB}}$ and their inverses lower and raise  projected indices. Since $\eta_{AB}$ and ${H}_{AB}$ are invariant under the action of $\hat{\cal L}$, $G$ and $\cal H$ we find, $\Gamma_{\overline A \underline B} = 0$, where $\Gamma_{AB}$ was defined in (\ref{Lambda}), and $$\Gamma_{\overline A \underline B} = \ov{P}_{A}{}^{C}P_{B}{}^{D}\Gamma_{CD}\, .$$

A crucial object for the consistency of the theory is the Lorentz covariant derivative. Acting on a generic vector this derivative is defined as
\be
\nabla_{oA}V_{B}= E_{oA}V_{B} + \omega_{oAB}{}^{C}V_{C}\, 
\ee
where $E_{oA}\equiv \sqrt2{E_o}_{A}{}^{M}{}\partial_{M}$ and $\omega_{oAB}{}^{C}$ is a spin connection that satisfies
\be
\omega_{oABC} = - \omega_{oACB}\, \qquad \mathrm{and} \qquad \omega_{oA\overline B \underline C} = \omega_{oA\underline B \overline C} = 0 \, ,
\ee
in order to be compatible with $\eta_{AB}$ and $H_{AB}$ respectively. 

Unlike general relativity, DFT consists of a generalized notion of geometry and there are not enough compatibility conditions to fully determine the generalized spin connection. Only the totally antisymmetric  and trace parts of $\omega_{oABC}$ can be determined in terms of $E_{oM}{}^{A}$  and $d_o$,  i.e.
\be
\omega_{o[ABC]} = - E_{o[A}E_{o}^{N}{}_{B}E_{oNC]} - \frac{\sqrt{2}}{3}f_{MNP}E_{o}^{M}{}_{A}E_{o}^{N}{}_{B}E_{o}^{P}{}_{C}\equiv -\frac13{F}_{oABC}\, ,
\label{gralspinconnectionE}
\ee
\be
\omega_{oBA}{}^{B} = - \sqrt{2}e^{2d_o}\partial_{M}\left(E_{o}^{M}{}_{A}e^{-2d_o}\right)  \equiv - F_{oA}\, ,
\label{gralspinconnectiontrace}
\ee
the latter arising from partial integration with the dilaton density. 

The ${\cal N}=1$ supersymmetric extension of DFT is achieved by adding a couple of generalized background spinor fields that act as supersymmetric partners of the bosonic fields: the generalized gravitino $\Psi_{o \ov{A}}$ and the generalized dilatino $\rho_{o}$. Under the action of the symmetry groups these fields behave as

\begin{center}
\begin{tabular}{||c c c c c||}
\hline
& $G$ & ${\cal H}_{L}$ & ${\cal H}_{R}$ & Diff \\ [0.5ex]
\hline\hline
$\Psi_{o \overline A}$ & $G$-invariant & ${\cal H}_{L}$-spinor & ${\cal H}_{R}$-vector & scalar($t=0$) \\
\hline
$\rho_{o}$ & $G$-invariant & ${\cal H}_{L}$-spinor & ${\cal H}_{R}$-invariant & scalar($t=0$) \\ [1ex]
\hline
\end{tabular}
\end{center}

The covariant derivative of spinor fields acquires an additional term in order to derive the spinor indices. For instance, the covariant derivative of the generalized background gravitino and generalized background dilatino are

\bea
\nabla_{oA}\Psi_{o\ov{B}} & = & E_{oA}\Psi_{o\ov{B}} + \omega_{oA\ov{B}}{}^{\ov{C}}\Psi_{o\ov{C}} - \frac{1}{4}\omega_{oA\un{BC}}\gamma^{\un{BC}}\Psi_{o\ov{B}}\, , \nn\\
\nabla_{oA}\rho_{o} & = & E_{oA}\rho_{o} - \frac{1}{4} \omega_{oA\un{BC}}\gamma^{\un{BC}}\rho_{o}\,  . 
\eea
The gamma matrices satisfy a Clifford algebra for $\underline {\cal H}$
\be
\left\{\gamma^{\un{A}},\gamma^{\un{B}}\right\} = - 2P_{o}^{\un{AB}}\ , \label{Cliff}
\ee
and we use the standard convention for antisymmetrization of $\gamma$-matrices $\gamma^{{\underline A \dots \underline B}}=\gamma^{[\underline{A}} \dots \gamma^{\underline{B}]}$.  

The generalized supersymmetry transformations of the fundamental fields are parameterized by an infinitesimal Majorana fermion $\epsilon$, that is a spinor of $O(1,9)_L$. These transformations can be written as 
\bea
\delta_{\epsilon}E_{oM}{}^{A} & = & \ov{\epsilon}\gamma^{[B}\Psi_{o}^{A]}E_{oMB}\, , \nn \\
\delta_{\epsilon} \Psi_{o\ov{A}} & = & \nabla_{o\ov{A}}\epsilon\, , \nn \\
\delta_{\epsilon}d_{o} & = & - \frac{1}{4}\ov{\epsilon}\rho_o\, , \nn \\
\delta_{\epsilon}\rho_{o} & = & - \gamma^{\un{A}}\nabla_{o\un{A}}\epsilon \, .
\eea
If we now include all the symmetries described in the previous subsection, the background fields transform as  
\bea
\delta E_{o}^{M}{}_{\un{A}}{} & = & \xi^{P}\partial_{P} E_{o}^{M}{}_{\un{A}}{} + (\partial^{M}\xi_{P} - \partial_{P}\xi^{M})E_{o}^{P}{}_{\un{A}}{} + E_{o}^{M}{}_{\un{B}}{}\Gamma^{\un{B}}{}_{\un{A}} - \frac12 \ov{\epsilon}\gamma_{\un{A}} \Psi_{o}^{\ov{B}}E_{o}^{M}{}_{\ov{B}}\, ,\nn \\   
\delta E_{o}^{M}{}_{\ov{A}}{} & = & \xi^{P}\partial_{P}E_{o}^{M}{}_{\ov{A}}{} + (\partial^{M}\xi_{P} - \partial_{P}\xi^{M})E_{o}^{P}{}_{\ov{A}}{} + E_{o}^{M}{}_{\ov{B}}{}\Gamma^{\ov{B}}{}_{\ov{A}} +\frac12 \ov{\epsilon}\gamma^{\un{B}}\Psi_{o\ov{A}}E_{o}^{M}{}_{\un{B}}{}\, ,\nn\\   
\delta d_o & = & \xi^{P}\partial_{P}d_o - \frac{1}{2}\partial_{P}\xi^{P} - \frac{1}{4}\ov{\epsilon}\rho_o \, , \label{0transf}\\   
\delta\Psi_{o\ov{A}} & = & \xi^{M}\partial_{M}\Psi_{o\ov{A}} + \Gamma^{\ov{B}}{}_{\ov{A}}\Psi_{o\ov{B}} + \frac{1}{4}\Gamma_{\un{BC}}\gamma^{\un{BC}}\Psi_{o\ov{A}} + \nabla_{o\ov{A}}\epsilon \, ,\nn\\
\delta\rho_o & = & \xi^{M}\partial_{M}\rho_o + \frac{1}{4}\Gamma_{\un{BC}}\gamma^{\un{BC}}\rho_o - \gamma^{\un{A}}\nabla_{o\un{A}}\epsilon \, .\nn
\eea
It is straightforward to show that the previous transformation close off-shell \footnote{In case of considering the full-order fermion transformations, the closure is given only on-shell.}
 with the following parameters 
\begin{align}
\begin{split}\label{par0}
\xi^{M}_{12} &= [\xi_1, \xi_{2} ]^{ M}_{(C_f)} - \frac{1}{\sqrt{2}}E_{o}^{M} {}_{\underline{A}}\overline{\epsilon_1} \gamma^{\underline{A}} \epsilon_{2} ,\\
\Gamma_{12 { A} { B}} &= 2 \xi_{[1}^{ P} \partial_{ P} \Gamma_{2] { A} { B}} - 2 \Gamma_{[1  A}{}^{ C} \Gamma_{2] {C B}}+E_{o[ A}\left(\overline\epsilon_1\gamma_{ B]}\epsilon_2\right)-\frac1{2}\left(\overline\epsilon_1\gamma^{\underline C}\epsilon_2\right)F_{o{AB}\underline C} , \\
\epsilon_{12} &= -\frac{1}{2} \Gamma_{[1 \underline{B} \underline{C}} \gamma^{\underline{B} \underline{C}} \epsilon_{2]} + 2 \xi_{[1}^{ P} \partial_{ P} \epsilon_{2]} \, ,
 \end{split}
\end{align}
where the $C_f$-bracket is defined as
\bea
 [\xi_1, \xi_{2} ]^{ M}_{(C_f)}=2\xi^{ P}_{[1}\partial_{ P}\xi_{2]}^{ M}-\xi_{[1}^{ N}\partial^{ M}\xi_{2] N}+f_{{ PQ}}{}^{ M} \xi_{1}^{ P} \xi_2^{ Q}\, .
\eea

The transformation rules of the background fields discussed in the previous subsection leave the following action invariant (up to leading order  terms in fermions)
\bea
\label{DFTsusyAction}
S_{\mathcal{N}=1} & = & \int d^{20}X e^{-2d_{o}} \Big( {\cal R}_{o} + L_{oF} \Big) \nn \\
& = & \int d^{20}X e^{-2d_{o}} \Big( {\cal R}_{o} + \ov{\Psi}_{o}^{\ov{A}}\gamma^{\un{B}}\nabla_{o\un{B}}\Psi_{o\ov{A}} - \ov{\rho}_{o}\gamma^{\un{A}}\nabla_{o\un{A}}\rho_{o} + 2\ov{\Psi}_{o}^{\ov{A}}\nabla_{o\ov{A}}\rho_{o} \Big) \, ,
\eea
where $L_{oF}$ is the fermionic part of the Lagrangian and ${\cal R}_{o}$ is the generalized Ricci scalar,
\bea
{\cal R}_{o} & = & 2E_{o\un{A}}F_{o}^{\un{A}} + F_{o\un{A}}F_{o}^{\un{A}} - \frac16{F}_{o\un{ABC}}F_{o}^{\un{ABC}} - \frac12{F}_{o\ov{A}\un{BC}}F_{o}^{\ov{A}\un{BC}}.\label{r}
\label{GRicci}
\eea
We can notice that the previous expression is written in terms of determined components of the generalized spin connection, even when it is obtained from a T-duality invariant curvature tensor ${\cal R}_{oABCD}$ which is not fully determined. Moreover, the covariant derivatives appearing in $L_{oF}$ are also fully determined and therefore the full ${\cal N}=1$ action is fully determined.   

The ${\cal N}=1$ DFT action is invariant under $G$, ${\cal H}$, generalized diffeomorphisms and supersymmetry. The  equations of motion obtained from \eqref{DFTsusyAction}, up to leading order terms in fermions, are
\bea
\label{EoMR}
{\cal R}_{o\underline{B} \overline A } + \bar{\Psi}_{o}^{\overline C} \gamma_{\underline B} E_{o \ov A} \Psi_{o \overline C} - \bar{\rho}_{o} \gamma_{\underline B} E_{o \overline A} \rho - 2 \bar{\Psi}_{o \overline A} E_{o \underline B} \rho_{o} & = & 0 \, , \nn \\ {\cal R}_{o} & = & 0 \, , \nn \\
\gamma^{\underline{B}}\nabla_{o\underline{B}}\Psi_{o\overline{A}}+\nabla_{o\overline{A}}\rho_{o} & = & 0 \, , \nn \\
\gamma^{\underline{A}} \nabla_{o\underline{A}}\rho_{o}+\nabla_{o\overline{A}}\Psi_{o}^{\overline{A}} & = & 0\, ,
\eea
where ${\cal R}_{o\underline{B} \overline A }$ is the bosonic part of the e.o.m of the generalized frame and the e.o.m of the fermionic fields have been used to simplified the equations.

Up to this point, we have described the basics of ${\cal N}=1$ DFT for generalized background fields. In the next section we perturb these background fields, asking for a linear perturbation of the generalized frame. This perturbation is compatible with ${\cal N}=1$ supersymmetry and reduces to a generalized Kerr-Schild ansatz when supersymmetry is turned off. Then we inspect how ${\cal N}=1$ supersymmetry is accomplished in the other fields of the theory.  

\section{The ${\cal N}=1$ supersymmetric generalized Kerr-Schild ansatz }
\label{GKS}
  
\subsection{Finite perturbations on the background fields}

We consider the most general linear perturbation for the generalized frame in the flux formalism of DFT. We start defining, 
\bea
E_{M}{}^{\ov A} = E_{oM}{}^{\ov A} + \frac12 \kappa E_{oM}{}^{\underline B} {\Delta}_{\underline B}{}^{\ov A}\, \nn \\ E_{M}{}^{\underline A} = E_{oM}{}^{\underline A} - \frac12 \kappa E_{oM}{}^{\ov B} {\Delta}^{\underline A}{}_{\ov B}{}
\label{GKSA}
\eea
with $\kappa$ an arbitrary parameter and $\Delta_{A}{}^{B}$ a mixed-projected perturbation that satisfies
\bea
\label{pcons}
\Delta_{\ov A \ov{B}} & = & 0\, , \\
\Delta_{\underline A \underline{B}} & = & 0 \, \nn, \\
\Delta \eta^{-1} \Delta & = & 0 \, , 
\label{pconstraints}
\eea
in order to be consistent with the constraints of DFT. There is no ambiguity in the contractions in (\ref{pconstraints}). The inclusion of a finite perturbation on the generalized background frame satisfying (\ref{pcons}) and (\ref{pconstraints})  only deforms the curved version of the projectors,     
\bea
P_{M N} & = & E_{M}{}^{\underline A} E_{N \underline A} = P_{o M N} - \kappa E_{o(M \overline A} \Delta_{\underline B}{}^{\ov A}{} E_{oN)}{}^{\underline B} \nn \\ \bar P_{M N} & = & E_{M}{}^{\ov A} E_{N \ov A} = \bar P_{o M N} + \kappa E_{o(M \underline A} \Delta^{\underline A}{}_{\overline B} E_{oN)}{}^{\overline B} \, \nn \\ P_{A B} & = & E_{M \underline A} E^{M}{}_{ \underline B} = P_{o A B}  \nn \\ \bar P_{\ov{A B}} & = & E_{M \ov A} E^{M}{}_{\ov B} = \bar P_{o \ov{A B}} \, . 
\eea
The ansatz (\ref{GKSA}) is compatible with ${\cal N}=1$ supersymmetry and reduces to the generalized Kerr-Schild ansatz introduced in \cite{KL} when one considers
\bea
\Delta_{\underline A \overline B} = K_{M} \bar{K}_{N} E^{M}{}_{\underline A} E^{N}{}_{\overline B} \, .
\label{DeltaK}
\eea
The perturbation $\Delta_{AB}$ is a $G$-singlet, ${\cal H}$-vector and a generalized scalar with weight $t=0$ with respect to generalized diffeomorphisms. The generalized background dilaton can be perturbed with a generic $\kappa$ expansion,
\be
d = d_{o} + \kappa f\, , \qquad f = \sum_{n=0}^{\infty}\kappa^{n}f_{n} \, ,
\ee
with $n\geq 0$. The function $f$ is a $G$-singlet, a ${\cal H}$-invariant and a scalar with weight $t=0$ under generalized diffeomorphisms. The previous expansion was introduced in \cite{KL} in the context of heterotic DFT but the expansion of the fermionic fields were not considered. As we are interested in this last point, we mimic the structure of the generalized perturbation of the generalized dilaton and propose
\be
\Psi_{\overline A} = \Psi_{o \overline A} + \kappa \Theta_{\overline A}\, , \qquad \Theta_{\overline A} = \sum _{n=0}^{\infty}\kappa^{n}\Theta_{n \overline A} \, ,
\ee
and
\be
\rho = \rho_{o} + \kappa g\, , \qquad g = \sum _{n=0}^{\infty}\kappa^{n}g_{n} \, .
\ee
Using the conventions of the previous section it is possible to find that $\Theta_{\overline A}$ is a $G$-singlet, a spinor of $O(9,1)_{L}$, a vector of $O(1,9+n)_{R}$ and a scalar with weight $t=0$ under generalized diffeomorphisms, and $g$ is a G-singlet, a spinor of $O(9,1)_{L}$, an invariant  of $O(1,9+n)_{R}$ and a scalar with weight $t=0$ under generalized diffeomorphisms.

In the next part of this work we explicitly show how supersymmetry truncates the $\kappa$ expansions for some of the the generalized background fields in order to be consistent with the supersymmetric extension of the generalized Kerr-Schild ansatz defined in (\ref{GKSA}).  

\subsection{Supersymmetric consistency conditions} \label{susyconditions}
We start analizing the supersymmetric transformation of $\Delta_{\underline A \overline B}$. Considering
\bea
\delta_{\epsilon} E_{M A} = \bar \epsilon \gamma^{[B} \Psi^{A]} E_{M B}
\eea
and proposing the $\kappa$ expansions discussed in the previous section we find,
\bea
\delta_{\epsilon} \Delta_{\underline A \overline B} = \bar \epsilon \gamma_{\underline A} \Theta_{\overline B} \, 
\label{susytransf}
,
\eea
where we have used 
\bea
\delta\cdots\delta (\Delta \eta^{-1} \Delta) = 0 , 
\label{deltan}
\eea
with $\delta$ a generic transformation. The expression (\ref{susytransf}) forces 
\bea
\Theta_{n} = 0 \quad , \quad n\geq1 \, .
\eea

On the other hand (\ref{susytransf}) is correct up to a generalized Lorentz transformation that can be reabsorbed in the generalized Lorentz parameter. Let us observe that the decomposition of $\Delta_{\underline A \overline B}$ in terms of null vectors $K_{M}$,$\bar{K}_{M}$ is not allowed since (\ref{susytransf}) cannot be solved for both vectors.

The supersymmetric transformation of $\Theta_{\overline A}$ is
\bea
\delta_{\epsilon} \Theta_{\overline A} = \frac12 \Delta^{\underline B}{}_{\overline A} E_{\underline B} \epsilon + \frac{1}{4\kappa} \tilde{F}_{\overline A \underline{BC}} \gamma^{\underline {BC}} \epsilon
\eea
where 
\bea
F_{\overline A \underline{BC}} = F_{o\overline A \underline{BC}} + {\tilde F}_{\overline A \underline{BC}} \, .
\eea
Since the perturbations on the fluxes are cubic in $\kappa$, we need to impose some supersymmetric consistency constraints on the generalized gravitino transformation. Explicitly we have,
\bea
\delta_{\epsilon} \Theta_{0 \overline A} = && \frac12 \Delta^{\underline B}{}_{\overline A} E_{\underline B} \epsilon + \frac{1}{4} \Big(\frac{1}{2} \Delta_{\underline D \overline A}  {F_o}^{\underline D}{}_{\underline{BC}} - \Delta_{\underline B}{}^{\overline D} F_{o\overline{AD}\underline C} + E_{\underline C} (\Delta_ {\underline B\overline A}) \nn \\ && + \frac{1}{\sqrt 2} f_{\underline{D B C}} \Delta^{\underline D}{}_{\overline A}  + \sqrt{2} f_{\overline A \underline C \overline D} \Delta_{\underline B}{}^{\overline D} \Big) \gamma^{\underline {BC}} \epsilon \, , 
\eea
\bea
\label{zer}
\delta_{\epsilon} \Theta_{1 \overline A} = && \Big( \frac{1}{4} \bar \Delta_{\underline B}{}^{\overline D}  \Delta_{\underline C}{}^{\overline E} F_{o\overline{ D E A}} - \frac{1}{2} F_{o \underline B \overline E}{}^{\underline D} \Delta_{\underline C}{}^{\overline E} \Delta_{\underline D \overline A} \\ && + \frac{1}{2} \Delta_{\underline B}{}^{\overline D} (E_{\overline D} \Delta_{\underline C \overline A})  + \frac{\sqrt 2 }{4} f_{\overline{ADE}} \Delta_{\underline B}{}^{\overline D} \Delta_{\underline C}{}^{\overline E} - \frac{1}{\sqrt 2} f_{\underline{D B} \overline E} \Delta^{\underline D}{}_{\overline A} \Delta_{\underline C}{}^{\overline E} \Big) \gamma^{\underline {BC}} \epsilon \nn \, ,
\eea

\be
\delta_{\epsilon} \Theta_{2 \overline A} = \Big( \frac{1}{8} \Delta_{\underline B}{}^{\overline E} \Delta_{\underline C}{}^{\overline F} \Delta_{\underline D \overline A} {F_o}^{\underline D}{}_{\overline{EF}} + \sqrt{2}  f_{\underline D \overline{EF}} \Delta^{\underline D}{}_{\overline A} \Delta_{\underline B}{}^{\overline E} \Delta_{\underline C}{}^{\overline F}  \Big) \gamma^{\underline {BC}} \epsilon \, , 
\label{zero}
\ee
where we have used the following notation $f_{ABC} = f_{MNP} E^{M}{}_{A} E^{N}{}_{B} E^{P}{}_{C}$. Therefore we impose the following supersymmetric consistency constraints, 
\bea
\delta\cdots\delta\left(\Theta_1\right) = \delta\cdots\delta\left(\Theta_2\right) = 0 \, ,
\label{scc}
\eea
in order to reproduce a linear $\kappa$ expansion for the generalized perturbed gravitino. This requirement cannot be solved invoking (\ref{deltan}) and thus (\ref{scc}) must be treated as extra constraints on the theory. The conditions (\ref{geodesicdft}) can be rewritten with the help of (\ref{DeltaK}) as
\bea
0 & = & F_{\underline A \overline{BC}} \Delta^{\underline A}{}_{\overline D} \Delta_{\underline B}{}^{\overline B} = F_{\overline A \underline{BC}} \Delta_{\underline D}{}^{\overline A} \Delta^{\underline B}{}_{\overline B} \, , \nn \\ 0 & = & f_{\underline A \overline{BC}} \Delta^{\underline A}{}_{\overline D} \Delta_{\underline B}{}^{\overline B} = f_{\overline A \underline{BC}} \Delta_{\underline D}{}^{\overline A} \Delta^{\underline B}{}_{\overline B} \, ,
\label{extrac1}
\eea
and therefore the remaining supersymmetric constraints are
\bea
0 = && \Big( \frac{1}{4} \bar \Delta_{\underline B}{}^{\overline D}  \Delta_{\underline C}{}^{\overline E} F_{o\overline{ D E A}}  + \frac{1}{2} \Delta_{\underline B}{}^{\overline D} (E_{\overline D} \Delta_{\underline C \overline A})  + \frac{\sqrt 2 }{4} f_{\overline{ADE}} \Delta_{\underline B}{}^{\overline D} \Delta_{\underline C}{}^{\overline E} \Big) \gamma^{\underline {BC}} \epsilon \, .
\label{extrac2}
\eea
By a similar argument we seek constraints in the generalized background dilatino transformation,  
\bea
\delta g = - \frac12 \gamma_{\underline A} \Delta^{\underline A}{}_{\overline B} E^{\overline B} \epsilon - \frac{1}{12\kappa} \tilde{F}_{\underline{ABC}} \gamma^{\underline{ABC}} \epsilon - \frac{1}{2\kappa} \tilde{F}_{\underline{B}} \gamma^{\underline B} \epsilon
\eea
where
\bea
F_{\underline A \underline{BC}} & = & F_{o\underline A \underline{BC}} + {\tilde F}_{\underline A \underline{BC}} \, \nn \\
{F}_{\underline{A}} & = & {F}_{o \underline A} + {\tilde F}_{\underline{A}} \, ,
\eea
and
\bea
\tilde{F}_{\underline {A B C}}= && - \frac{3\kappa}{2}(\Delta_{[\underline A}{}^{\overline D} F_{0\overline D|\underline{BC}]} + \sqrt{2} f_{\overline D[\underline{AB}} \Delta_{\underline C]}{}^{\overline D}) \nn \\ && + \frac{3 \kappa^2}{4} (\Delta_{[\underline A}{}^{\overline D} \Delta_{\underline B}{}^{\overline E} F_{0 \overline{DE}|\underline{C}]}  + \sqrt{2} f_{[\underline A|\overline{DE}} \Delta_{\underline B}{}^{\overline D} \Delta_{\underline C]}{}^{\overline E}) \nn \\ && - \frac{3 \kappa^3}{8} (\Delta_{\underline A}{}^{\overline D} \Delta_{\underline B}{}^{\overline E} \Delta_{\underline C}{}^{\overline F} F_{0 \overline{DEF}} + \sqrt{2} f_{\overline{DEF}} \Delta_{\underline A}{}^{\overline D} \Delta_{\underline B}{}^{\overline E} \Delta_{\underline C}{}^{\overline F}) \nn \, ,
\eea

\bea
{\tilde F}_{\underline{B}} = - 2 E_{0\underline B} f + \frac{\kappa}{2} (\Delta_{\underline B}{}^{\overline C} \omega_{0\underline A \overline C}{}^{\underline A} + E_{0 \overline C} \Delta_{\underline B}{}^{\overline C}) \, .
\eea
Because of the appearance of $f$ in the last expression, we have an infinite $\kappa$ expansion for the generalized dilatino that can be solved once the generalized dilaton is solved. The previous statement means that the $\kappa$ expansion of these fields are not restricted by supersymmetry.

\subsection{Perturbed action and equations of motion}
Up to this point, we have perturbed the field content of ${\cal N}=1$ DFT in a consistent way. The action of the perturbed theory must be of the same form as (\ref{DFTsusyAction}), \textit{i.e.}
\be
S_{\mathcal{N}=1} = \int d^{20}X e^{-2d_{}}{\cal R}_{} + \ov{\Psi}_{}^{\ov{A}}\gamma^{\un{B}}\nabla_{\un{B}}\Psi_{\ov{A}} - \ov{\rho}_{}\gamma^{\un{A}}\nabla_{\un{A}}\rho_{} + 2\ov{\Psi}_{}^{\ov{A}}\nabla_{\ov{A}}\rho_{}\, .
\ee
and the equations of motion up to leading order terms in fermions, are
\bea
{\cal R}_{\underline{B} \overline A } + \bar{\Psi}^{\overline C} \gamma_{\underline B} E_{\ov A} \Psi_{\overline C} - \bar{\rho} \gamma_{\underline B} E_{\overline A} \rho - 2 \bar{\Psi}_{\overline A} E_{\underline B} \rho & = & 0, \nn \\ {\cal R}_{} & = & 0, \, \nn \\
\gamma^{\underline{B}}\nabla_{\underline{B}}\Psi_{\overline{A}}+\nabla_{\overline{A}}\rho_{} & = & 0, \nn \\
\gamma^{\underline{A}} \nabla_{\underline{A}}\rho_{}+\nabla_{\overline{A}}\Psi_{}^{\overline{A}} & = & 0\, .
\eea 
 
Since the generalized geodesic equations introduced in (\ref{geodesicdft}) cannot be defined in terms of $\Delta_{AB}$, ${\cal R}_{\underline B \overline A}$ has cubic contributions of the perturbation parameter $\kappa$ coming from the generalized fluxes \footnote{Higher order terms are identically null.}. As a consequence, the generalized equations of motions are no longer quadratic in $\kappa$ even if $f=g=0$, unlike the result obtained in \cite{KL}.

In the next section we proceed to parametrize the previous field content and find the necessary conditions to obtain the ${\cal N}=1$ supersymmetric extension of the ordinary Kerr-Schild ansatz in the context of the low energy effective heterotic field theory. We start reviewing the parametrization of the background field content and then we go straightforwardly to the perturbative theory.

\section{Reduction to ${\cal N}=1$ supergravity}
\label{Param}

\subsection{Parameterization of the background field content}

We start by splitting the $G$ and $\cal H$  indices  as ${ M}=({}_\mu,{}^\mu,i)$ and ${ A}=(\underline{A},\overline{A})$ with $\underline{A}=\underline a, \overline{A}=(\overline{a},\overline{i})$,  respectively,  $_\mu, ^\mu, \underline a, \overline a=0,\dots, 9$, $i, \overline i=1,\dots, n$. The parametrization of the fundamental background fields of ${\cal N}=1$ DFT must respect all the constraints of the theory. The generalized background frame is an $O(d,d+n)$ element, so it is parametrized in the following way,
\be
E_{o}^{M}{}_{ A}  =\left(\begin{matrix}{ E}_{o\mu \underline a}&  { E}_{o}^{\mu }{}_{\underline a} & E_{o}^i{}_{\underline a}\\ 
E_{o\mu \overline  a}& E_{o}^\mu{}_{\overline  a}&E_{o}^i {}_{\overline a} \\
E_{o\mu\overline i} &E_{o}^\mu{}_{\overline i} &E_{o}^i{}_{\overline i} \end{matrix}\right) \ = \
\frac{1}{\sqrt{2}}\left(\begin{matrix}-{ e}_{o\mu a}-C_{o \rho\mu} { e}_{o}^{\rho }{}_{a} &  { e}_{o}^{\mu }{}_{a} & -A_{o\rho}{}^i { e}_{o}^{\rho }{}_{{a}}\, , \\ 
\overline e_{o\mu a}-C_{o\rho \mu}{} \overline e_{o}^{\rho }{}_{{a}}& \overline e_{o}^\mu{}_{a}&-A_{o\rho}{}^i  \overline e_{o}^\rho{}_{a} \\
\sqrt2 A_{o\mu i}e^i{}_{\overline i} &0&\sqrt2 e^i{}_{\overline i} \end{matrix}\right)  \, ,
\label{HKparam}
\ee
where ${e}_{o \mu a}$ and $\overline e_{o\mu a}$ satisfy
\bea
e_{o\mu a} \eta^{ab} e_{o\nu b} = \overline e_{o\mu a} \eta^{ab} \overline e_{o\nu b} = g_{o\mu \nu} \, , 
\label{gf}
\eea
with $\eta_{ab}$ the ten dimensional flat metric, $a,b=0,\dots, 9$,  $C_{o\mu\nu}=b_{o\mu\nu}+\frac12 A_{o\mu}^i A_{o\nu i}$, with $A_{o \mu}^i$ being the gauge connection. The invariant projectors of DFT are parametrized in the following way
\bea
P_{\underline{ab}}=-\eta_{{ab}}\delta_{\underline a}^a\delta_{\underline b}^b, \quad \overline P_{\overline{ab}}=\eta_{{ab}}\delta_{\overline a}^a\delta_{\overline b}^b, \quad \overline P_{\overline{ij}}=e_{\ov{i}}{}^{i} \kappa_{ij} e_{\ov{j}}{}^{j} = \kappa_{\ov{i j}}\,\, , 
\eea 
where $\kappa_{ij}$ and $\kappa_{\bar i \bar j}$ are the Cartan-Killing metrics associated with the $SO(32)$ or $E_8\times E_8$ heterotic gauge group. The gauge fixing (\ref{gf}) imposes
\bea
\delta e_{o\mu a} = \delta \overline e_{o\mu a},
\label{gaugef}
\eea
and therefore the parametrization of the components of the generalized Lorentz parameters are not independent
\bea
 \Gamma_{\underline{ab}} \delta^{\underline{ab}}_{ab} & = & \left( -\Lambda_{{ab}}  + \bar{\epsilon} \gamma_{[a} \psi_{ob]}\right) \, ,\nn \\   
\Gamma_{\overline{ab}} \delta^{\ov{ab}}_{ab} & = & \Lambda_{{ab}} \, ,
\label{lambdagaugefixing}
\eea 
where $\Lambda_{ab}$ denotes  the generator of the  $O(1,9)$ transformations. We also impose $\delta E^i{}_{\overline i}=0$ and $\delta E^\mu{}_{\overline i}=0$ which leads to 
\bea
\Gamma_{\overline{ij}}= f_{{ijk}}\xi^{ k} \delta_{\overline i}^i\delta_{\overline j}^j\, \qquad {\rm and}\qquad \Gamma_{\overline {ai}}=-\Gamma_{\overline{ ia}}=\frac1{2\sqrt2} \overline\epsilon\gamma_a \chi_{o i}\ \delta^a_{\overline a}\ \delta^i_{\overline i}\, , \label{lorgau}
\eea
where we have parameterized the generalized gravitino field as
\bea
\Psi_{{oA}} & = & (0, e_{o}^\mu{}_{ a}\psi_{o\mu},\frac1{\sqrt2}e^i{}_{\overline i}\chi_{oi} ) \, .
\eea

The structure constants are trivially incorporated,
\bea
\label{fidentification}
f_{{MN}}{}^{P}=\left\{\begin{matrix}f_{ij}{}^k & {\rm for} \quad { {M}, {N}, {P}}=i,j,k \\
0 & {\rm otherwise.}
\end{matrix}\right.
\eea
In addition we parameterize
\bea
\xi^{ M}=(\xi^\mu, \lambda_\mu, \xi^i) \, ,
\eea
where the parameter $\xi^{\mu}$ is associated with the usual Lie derivative, defined as
\bea
{\cal L}_{\xi} v^{\mu} = \xi^{\nu} \partial_{\nu} v^{\mu} + (\partial_{\nu} \xi^{\mu}) v^{\nu} \, ,
\eea
with $v^{\mu}$ a generic vector. The parameter $\lambda_{\mu}$ parameterizes the abelian gauge symmetry of the background Kalb-Ramond field,
\bea
\delta_{\lambda} b_{o\mu\nu}=2\partial_{[\mu}\lambda_{\nu]} \, ,
\eea
while $\xi_i$ is the non-abelian gauge parameter. On the other hand, the parametrizations of the generalized background dilaton and dilatino are
\bea
d & = & \phi_o-\frac{1}{2}\log\sqrt{-g_o} \, , \nn \\ \rho & = &2\lambda_{o} +\gamma^a \psi_{oa} \, .
\eea

The $\gamma$-functions  $\gamma^{\underline a}=\gamma^a\delta_a^{\underline a}$ verify the Clifford algebra
\bea
\{\gamma^a, \gamma^b\}=2\eta^{ab}
\eea 
and  the supersymmetric transformation rules of the background field content are 
\bea
\delta_\epsilon e_{o\mu}{}^{a} &= & \frac12 \bar{\epsilon} \gamma^{a} \psi_{o \mu}\, , \  \qquad   \qquad \qquad \ \ \   \delta_\epsilon \psi_{o \mu} = \partial_{\mu} \epsilon -\frac14w^{(+)}_{o\mu ab}\gamma^{ab}\epsilon \nn \,  ,  
 \\
\delta_\epsilon b_{o\mu \nu} & = & \bar{\epsilon} \gamma_{[\mu} \psi_{o \nu ] }+\frac12\bar\epsilon\gamma_{[\mu}\chi_{o}^i A_{o\nu]i}\, , \ \   
    \delta_\epsilon \lambda_{o} \ = \ - \frac{1}{2}\gamma^{a} \partial_{oa}\phi_{o} \epsilon +\frac{1}{24}H_{o{a} {b} c}\gamma^{{a} {b} c} \epsilon \, , \nn \ \    \\
\delta_\epsilon \phi_{o} &=&  -  \frac{1}{2} \bar{\epsilon} \lambda_{o} \, , \qquad\qquad  \qquad \
\delta_\epsilon A_{o\mu}^i = \frac12\bar\epsilon\gamma_{\mu} \chi_{o}^i\, ,  \qquad  \delta_\epsilon\chi_{o}^i=-\frac14 F_{o \mu\nu}^i\gamma^{\mu\nu}\epsilon 
\label{susysugra} 
\eea
where 
\bea
w^{(\pm)}_{o\mu ab}&=&-e_{o}^\mu{}_{[a}{} e_{o}^\nu{}_{b]}{}\partial_\mu e_{o\nu c} +e_{o}^\mu{}_{[a}{} e_{o}^\nu{}_{c]}{}\partial_\mu e_{o\nu b}
+e_{o}^\mu{}_{[b}{} e_{o}^\nu{}_{c]}{}\partial_\mu e_{o\nu a}\pm \frac12 H_{o\mu\nu\rho} e_{o}^\nu{}_a e_{o}^\rho{}_b\, ,\nn\\
F_{o \mu\nu}{}^{i}&=&2\partial_{[\mu}A_{o\nu]}{}^{i}-f^i{}_{jk}A_{o\mu}^jA_\nu^k \, , \nn  \\ H_{oabc}&=& =3e_{oa}^\mu e_{ob}^\nu e_{oc}^\rho\left(\partial_{[\mu}b_{o\nu\rho]}-A_{o[\mu}^i\partial_\nu A_{o\rho]i}+\frac13
f_{ijk}A_{o \mu}^i A_{o\nu}^j A_{o\rho}^k\right) \nn \, .
\eea

The transformations (\ref{susysugra}) leave the low energy effective heterotic action invariant
\bea
S_{o} &=& \int d^{10}x \ e_{o} \ e^{-2\phi_{o}}\left[R_{o} -\frac{1}{12}H_{o \mu \nu \rho}H_{o}^{\mu \nu \rho} + 4\partial_{\mu}\phi_{o} \partial^{\mu}\phi_{o}-\frac14 \textrm{tr}( F_{o\mu\nu}F_{o}^{\mu\nu}) \right. \nn\\ 
&  & \ \ \ \ \ \ \ \ \ \ \ \ \ \ \ \ \ \  - \bar{\psi}_{o \mu}\gamma^{\mu \nu \rho}{D}_{\nu}\psi_{o \rho}  + 4\bar{\lambda}_{o}\gamma^{\mu \nu}{D}_{\mu}\psi_{o \nu} + 4\bar{\lambda}_{o} \gamma^{\mu}{D}_{\mu}\lambda_{o} -\frac12 \textrm{tr}(\bar\chi_{o}{\slashed{ D}}\chi_{o})\nn\\
&&\ \ \ \ \ \ \ \ \ \ \ \ \ \ \ \  \ \  + \ 4\bar{\psi}_{o \mu}\gamma^{\nu}\gamma^{\mu}\lambda \partial_{\nu}\phi_{o} - 2\bar{\psi}_{o \mu}\gamma^{\mu}\psi_{o}^{\nu}\partial_{\nu}\phi_{o} -\frac14\bar \chi_{o i} \gamma^\mu\gamma^{\nu\rho}F_{o \nu\rho}^i\left(\psi_{o \mu}+\frac13\gamma_\mu\lambda_{o}\right)\nn \\
& & \left. \ \ \ \ \ \ \ \ \ \ \ \ \ \ \ \ \ \  + \frac{1}{24}H_{o\rho \sigma \tau}\left(\bar{\psi}_{o \mu}\gamma^{[\mu}\gamma^{\rho \sigma \tau}\gamma^{\nu ]}\psi_{o \nu} + 4\bar{\psi}_{o \mu}\gamma^{\mu \rho \sigma \tau}\lambda_{o} - 4\bar{\lambda}_{o} \gamma^{\rho \sigma \tau}\lambda_{o} + \frac12\bar\chi_o^i \gamma^{\rho\sigma\tau} \chi_{o i}\right) \right] \, .\nn\\
\label{BdRAction}
\eea

The conventions for the  Riemann tensor are
\bea
R_{o}^\rho{}_{\sigma\mu\nu}=e_{o}^{\rho a}e_{o\sigma }{}^bR_{o \mu\nu ab}&=&e_{o}^{\rho a} e_{o\sigma }{}^b\left(-2\partial_{[\mu} w_{o\nu]ab} + 2w_{o[\mu| a}{}^{{c}} w_{o|\nu] cb}\right)
\, ,
\eea
and therefore the Ricci scalar is
\bea
R_{o}=R_{o \mu\nu}{}^{ab}e_{o}^\mu{}_a e_{o}^\nu{}_b \, .
\eea

\subsection{Parameterization of the perturbations}

In section \eqref{GKS} we introduce the supersymmetric extension of the generalized Kerr-Schild ansatz in the flux formalism of DFT. Now we proceed with the parametrization of the perturbations of the generalized fields. 

We start by considering that both components of the generalized frame
\bea
E_{M\overline A} & = &  E_{oM\overline A} + \frac{\kappa}{2} \Delta_{\underline B \ov A} E_{0M}{}^{\underline B} \nn \\ E_{M\underline A} & = &  E_{oM\underline A} - \frac{\kappa}{2} \Delta_{\underline A \ov B} E_{0M}{}^{\overline B}
\eea
are $O(10,10+n)$ elements. So we can parametrize them as
\be
E^{M}{}_{ A}  =\left(\begin{matrix}{ E}_{\mu \underline a}&  { E}_{}^{\mu }{}_{\underline a} & E_{}^i{}_{\underline a}\\ 
E_{\mu \overline  a}& E_{}^\mu{}_{\overline  a}&E_{}^i {}_{\overline a} \\
E_{\mu\overline i} &E_{}^\mu{}_{\overline i} &E_{}^i{}_{\overline i} \end{matrix}\right) \ = \
\frac{1}{\sqrt{2}}\left(\begin{matrix}-{ e}_{\mu a}-C_{\rho\mu} { e}_{}^{\rho }{}_{a} &  { e}_{}^{\mu }{}_{a} & -A_{\rho}{}^i { e}_{}^{\rho }{}_{{a}}\, , \\ 
\overline e_{\mu a}-C_{\rho \mu}{} \overline e_{}^{\rho }{}_{{a}}& \overline e_{}^\mu{}_{a}&-A_{\rho}{}^i  \overline e_{}^\rho{}_{a} \\
\sqrt2 A_{\mu i}e^i{}_{\overline i} &0&\sqrt2 e^i{}_{\overline i} \end{matrix}\right)  \, 
\label{HKparam1}
\ee
where ${e}_{\mu a}$ and $\overline e_{\mu a}$ satisfy
\bea
e_{\mu a}\eta^{ab} e_{\nu b} = \overline e_{\mu a} \eta^{ab} \overline e_{\nu b} = g_{\mu \nu} \, . 
\eea
Condition (\ref{gaugef}) forces
\bea
\Delta_{\underline A \overline B} = (\Delta_{ab} \delta^{ab}_{\underline a \overline b},\Delta_{ai} \delta^{ai}_{\underline a \overline i}) \, ,
\eea
where $\Delta_{ab}$ is a symmetric perturbation that verifies
\bea
\Delta_{ab} g^{bd} \Delta_{cd} + \Delta_{ai} \kappa^{ij} \Delta_{cj}  = 0 \, ,
\eea
and
\bea
\Delta_{ai} g^{ab} \Delta_{bj} & = & 0 \nn \, , \\
\Delta_{ai} g^{ab} \Delta_{bc} & = & 0 \, , \nn \\ \Delta_{ai} g^{ab} \Delta_{bj} & = & 0 \, .
\eea
The previous parametrization can be decomposed in the following way,
\bea
\label{odecomp}
\Delta_{ab} & = & (\frac{1}{1+\frac12 \kappa l.\bar{l}}) l_{a} \bar{l}_{b} \, , \\ \Delta_{ai} & = & (\frac{1}{1+\frac12 \kappa l.\bar{l}}) l_{a} j_{i} \, ,
\label{decomp}
\eea
where $l_{a} = e^{\mu}{}_{a} l_{\mu}$ is the rotation of the null vector associated to the perturbation of the metric \eqref{sugrametric}, that satisfies
\bea
l_{a} \eta^{ab} l_{b} = 0 \, ,
\eea
and $\bar{l}_{a}$ is an auxiliary vector that satisfies a relaxed null condition,
\bea
\bar{l}_{a} \eta^{ab} \bar{l}_{b} + j_{i} \kappa^{i j} j_{j} = 0 \, .
\label{Relaxed}
\eea

A very interesting aspect of (\ref{decomp}) is that the supersymmetric extension of the Kerr-Schild formalism can be done in terms of a pair of vectors, as we are going to verify. Using (\ref{HKparam1}) and recalling that the generalized frame is an element of $O(10,10+n)$ it is straightforward to find,
\bea
g_{\mu \nu} & = & g_{o \mu \nu} + \frac{\kappa}{1+\frac12 \kappa l.\bar{l}} l_{(\mu} \bar{l}_{\nu)} \nn \\ b_{\mu \nu} & = & b_{o \mu \nu} - \frac{\kappa}{1+\frac12 \kappa l.\bar{l}} l_{[\mu} \Big( \bar{l}_{\nu]} - \frac{1}{\sqrt{2}} A_{\nu]}{}^{i} j_{i} \Big) \nn \\ A_{\mu i} & = & A_{o \mu i} + \frac{1}{\sqrt{2}} \frac{\kappa}{1+\frac12 \kappa l.\bar{l}} l_{\mu} j_{i} \, .
\label{bosonicp}
\eea
From the previous expression we note that the standard Kerr-Schild ansatz can be obtained in the case $l_{a}=\bar{l}_{a}$. On the other hand, the perturbation of the $10$-dimensional gravitino is 
\bea
\psi_{a} = \psi_{o a} + \kappa \Theta_{0 a} \, .
\eea

The supersymmetric transformation of $l_{a}$ and $\bar{l}_{a}$ in terms of $\Delta_{ab}$ can be read from (\ref{susytransf}). When we parametrize it we find, 
\bea
\delta_{\epsilon} \Delta_{ab} = \bar{\epsilon} \gamma_{a} \Theta_{0 b} + \frac12 \Delta^{c}{}_{b} \bar{\epsilon} \gamma_{a} \psi_{o c}   \, ,
\eea
where the second term comes from the gauge fixing (\ref{lambdagaugefixing}) of the double Lorentz parameters. In this point we identify
\bea
\Theta_{0 a} & = & \frac{-1}{4(1+\frac12 \kappa l.\bar{l})} ( \bar{l}_{a} l_{b} + \bar{l}_{b} l_{a}  ) \psi_{o}^{b} \, ,
\label{extra1}
\eea
to finally obtain
\bea
\delta_{\epsilon} \Delta_{ab} = 0 \, .
\label{susyreq}
\eea
This requirement allows to decompose $\Delta_{ab}$ in terms of a pair of vectors. Moreover the transformation rule of the latter is,
\bea
\delta l_{a} & = & \xi^{\mu} \partial_{\mu} l_{a} + l_{b} \Lambda^{b}{}_{a} \label{t1} \\ \delta \bar{l}_{a} & = & \xi^{\mu} \partial_{\mu} \bar{l}_{a} + \bar{l}_{b} \Lambda^{b}{}_{a} .
\label{t2}
\eea
In the previous expressions we recognise a scalar transformation with respect to diffeomorphisms and a local Lorentz transformation. It is important to stress that (\ref{t1}) and (\ref{t2}) do not receive a supersymmetric transformation since both vectors are on equal foot. Moreover if both of them receive a supersymmetric transformation, then we would not be able to explicitly write $\delta_{\epsilon} l(l)$ and $\delta_{\epsilon \bar l}(\bar l)$, and then we were forced to work with a $\Delta_{a b}$ perturbation as happens in the DFT scheme. In consequence we demand (\ref{susyreq}) in order to recover the supersymmetric extension of the parametrization of the generalized Kerr-Schild ansatz in terms of $l$ and $\bar l$.   

As we discussed in the previous section, the perturbation of the dilaton and dilatino are not constrained by supersymmetry,   
\bea
\phi & = & \phi_{o} + \kappa f \, , \nn \\ 
\lambda & = & \lambda_{o} + \frac{\kappa}{2} g \, .
\eea
However here we remark that these fields cannot be perturbed separately or using different orders in $\kappa$ for each perturbation. Finally using $\delta_{\epsilon} j_{i} = 0$, the perturbation of the gaugino is constrained in the following way
\bea
\chi_{i} = \chi_{o i} - \frac{\kappa}{2} l_{b} j_{i} \psi^{b}_{o} \, . 
\label{chip}
\eea

\section{Application of the model}
\label{app}
\subsection{Supergravity solutions}

The generalized Kerr-Schild ansatz is a powerful tool which consists in an exact perturbation of a background metric tensor $g_{o\mu \nu}$, a background gauge field $A_{o \mu i}$, a background antisymmetric tensor $b_{o \mu \nu}$ and a background scalar field $\phi_{o}$ using a pair of null vectors  $l_{a}$ and $\bar{l}_{a}$ \footnote{In Kerr-Schild heterotic supergravity $\bar{l}$ is a relaxed null vector \textit{cf.} (\ref{Relaxed}).}, that coincides with the ordinary Kerr-Schild ansatz when
\bea
l_{a}=\bar{l}_{a} \, . 
\eea
As discussed before this scheme can be written in terms of multiplets of $O(d,d+n)$ doubling the coordinates of the space and imposing the strong constraint. Here we present the ${\cal N}=1$ supersymmetric extension of this formalism considering $d=10$. The latter provides a formulation that can be solved to extend supergravity solutions. 

The ${\cal N}=1$ supersymmetric extension of a generic supergravity solution consist in an exact perturbation of the supersymmetric degrees of freedom. For instance, a generic perturbation of a gravitino field is
\bea
\psi_{a} & = & \psi_{o a} + \kappa {\Theta}_{o a} \, ,
\eea
where ${\Theta}_{a}$ is a generic spinorial proposal. In principle, the compactification of these kind of solutions in a $T^{k}$ cannot be written in terms of $O(k,k)$ multiplets \cite{Mariana} and therefore the DFT rewriting cannot be performed in arbitrary circumstances. If we are interested in obtaining such rearrangement of the field content, then $\Theta_{a}$ must be related with $l$ and $\bar{l}$ through
\bea
\Theta_{0 a} & = & \frac{-1}{4(1+\frac12 \kappa l.\bar{l})} ( \bar{l}_{a} l_{b} + \bar{l}_{b} l_{a}  ) \psi_{o}^{b} \, ,
\eea
as we computed in the previous section. In other words, the perturbations of a supergravity model can be strongly constrained by T-duality before compactification, and DFT provides a systematic method to obtain these constraints.

\subsection{Gaugino condensation}

It is well known that Poincar\'e invariance requires the expectation values of the individual fermions to vanish, which
does not need to be the case for fermion bilinears such as
\bea
\Sigma_{\mu \nu \rho} = c \, \textrm{tr} \Big( \bar{\chi} \gamma_{\mu \nu \rho} \chi \Big) \, .
\label{gamma3}
\eea
with $c$ a constant. These kind of terms can be considered as a deformation of the $H$-flux in heterotic supergravity. The gaugino bilinear (\ref{gamma3}) does not change the Bianchi identity of the $H$-flux up to order $\a'$ \footnote{Note that in this work we have used units such that $\ap=1$.} \cite{Gaugino},
\bea
dH - \frac{1}{4} \Big(\textrm{tr}(R^-)^2 - \textrm{tr}(F^2) \Big) = 0 \, ,
\eea
where $R^-$ is a Riemann tensor constructed with the torsionful spin-connection \cite{BdR}. With a non-trivial gaugino bilinear the natural definition for the $H$-flux is,
\bea
\tilde{H}_{\mu \nu \rho} = H_{\mu \nu \rho} + \Sigma_{\mu \nu \rho} \, .  
\eea
A generic Kerr-Schild ansatz for these kind of models admits perturbations in the gaugino field. However, if we are interested in rewriting the theory using DFT, the gaugino cannot be perturbed
\bea
\Theta_{o i} = 0 \, ,
\label{chipert}
\eea
since duality forces (\ref{chip}). Equation (\ref{chipert}) partially simplifies the supersymmetric contributions to the action and equations of motion of the fields. 

The dynamics of the fermionic sector of the heterotic supergravity when the generalized Kerr-Schild ansatz is considered, is dictated by the equations of motion which can be obtained from (\ref{BdRAction}) as we show in appendix \ref{appendix}. We work with background gaugino condensation, \textit{i.e.} $\chi_{i}=\chi_{o i}$  when bilinears of this field appears and we consider exact perturbations in the bosonic degrees of freedom. 

Let us start by considering the $b_{\mu \nu}$ e.o.m admitting gaugino condensation,
\bea
\Delta b_{\mu \nu} & = & - D^{\rho}\phi_{o} H_{\rho\mu\nu} + \frac{1}{2}D^{\rho}H_{\rho\mu\nu} + \frac{1}{4} D^{\rho}\phi \left(\frac{1}{2}\bar{\chi}_{o}^{i}\gamma_{\rho\mu\nu}\chi_{o i}\right)\, \nn \\
& & - \frac{1}{8}D^{\rho}\left(\frac{1}{2}\bar{\chi}_{o}^{i}\gamma_{\rho\mu\nu}\chi_{o i}\right) \, .
\eea
In view of (\ref{gamma3}) we set $c=-\frac18$ and the 3-form now is defined as
\bea
{\tilde H}_{\mu \nu \rho} = H_{\mu \nu \rho} - \frac18 \textrm{tr} \Big( \bar{\chi}_{o} \gamma_{\mu \nu \rho} \chi_{o} \Big) \, . 
\label{Htilde}
\eea
Next we write the e.o.m of the dilaton, gauge field and metric in terms of the curvature (\ref{Htilde}), 
\bea
\Delta \phi & = & R -\frac{1}{12} \tilde{H}_{ \mu \nu \rho} \tilde{H}^{\mu \nu \rho} + 4\partial_{\mu}\phi_{} \partial^{\mu}\phi -\frac14 \textrm{tr}( F_{\mu\nu}F^{\mu\nu}) \, , \nn \\
\Delta A_{\mu}{}^{i} & = & A_{\rho}{}^{i}\Delta b^{\rho}{}_{\mu} + \frac{1}{2}\tilde{H}_{\mu}{}^{\nu\rho}{F}_{\nu\rho}{}^{i} - 2D^{\nu}\phi {F}_{\nu\mu}{}^{i} + D^{\nu}{ F}_{\nu\mu}{}^{i} \, , \nn \\ 
\Delta g_{\mu \nu} & = & R_{\mu\nu} + 4D_{\mu}\phi D_{\nu}\phi - \frac{1}{4}\tilde{H}_{\mu\lambda\rho} \tilde{H}_{\nu}{}^{\lambda\rho} - \frac{1}{2} {F}_{\mu\lambda i} {F}_{\nu}{}^{\lambda i} \, . 
\label{eomtilde}
\eea
We stress that other bilinear combinations also preserve Lorentz invariance, making possible different kind of fermionic condensation in heterotic supergravity. In the next section we explore the supersymmetric extension of the fundamental charged heterotic string in $d=10$. In this solution the gauge field mimics a generalization of the Coulomb potential, and a non-trivial gaugino condensation is the most simple supersymmetric extension to the formalism.  

\subsection{Fundamental charged heterotic string}
In order to include the gaugino condensation in a particular generalized Kerr-Schild solution, let us elaborate on the ${\cal N}=1$ supersymmetric extension of the fundamental charged heterotic string solution \cite{Black},   
\bea
ds^{2} = \frac{1}{1+NH(r)}(-dt^{2}+ (dx^{9})^2) + \frac{q^2 H(r)}{4N(1+NH(r))^2}(dt+dx^{9})^2 + \sum_{i=1}^{8} dx^{i} dx^{i} \, , 
\eea
where $H(r)$ is a Green function and $N$ is a constant. The non-vanishing components of the two form and gauge field are
\bea
b_{9t} & = &\frac{NH(r)}{1+NH(r)} \,, \\
A^1_{0} & = & A^{1}_{9} = \frac{qH(r)}{1+NH(r)} \, ,
\eea
with $q$ a charge and $\phi=-\frac12 \textrm{ln}(1+N H(r))$. If we want to write this solution in its generalized Kerr-Schild form we need to introduce the $\varphi$ function in the ansatz redefining $\kappa_{DFT} \rightarrow \kappa \varphi$. Then we identify,
\bea
l & = & dt+dx^{9} \, , \nn \\
\bar{l} & = & -dt + \frac{4N^2-q^2}{4N^2+q^2} dx^{9} \, , \nn \\ j_{1} & = & \frac{4qN}{4N^{2} + q^2}
\eea
and
\bea
\kappa \varphi = \frac{(4N^2 + q^2)H}{4N} \, \label{charge}.
\eea
The identification (\ref{charge}) is related with the effective charge of a higher-dimensional generalization of the Coulomb potential for the single copy of this solution \cite{KL}. 

We can consider a fermionic condensation for this setup computing the curvature of the 2-form and including the contribution (\ref{gamma3}) for this model,
\bea
\tilde{H}_{it9} & = & \partial_{i} b_{t9} - \frac18 \textrm{tr}(\bar{\chi}_{o} \gamma_{it9} \chi_{o}) \, .
\eea
Dynamics for this geometry is dictated by equations (\ref{eomtilde}) and the ${\cal N}=1$ supersymmetric DFT rewriting is possible in this scenario, where the background generalized metric is the only generalized perturbed field, since the generalized dilaton remains unperturbed. 

\section{Conclusions}
\label{Conclu}

In this work we present the supersymmetric extension of the Generalized Kerr-Schild ansatz in the flux formulation of ${\cal N}=1$ supersymmetric DFT. This ansatz is compatible with ${\cal N}=1$ supersymmetry as long as it is not written in terms of generalized null vectors. We find that imposing a set of supersymmetric consistency conditions the perturbation of the generalized gravitino is linear in $\kappa$. The perturbations of the generalized dilaton and dilatino have no restrictions. 

When we parametrize the theory in terms of the field content of the low energy effective heterotic supergravity, we find linear perturbations for the 10-dimensional vielbein, Kalb-Ramond field, gauge field, gravitino and gaugino in terms of a pair of vectors and an unrestricted perturbation for the 10-dimensional dilaton and dilatino. Moreover, the supersymmetric conditions found in the ${\cal N}=1$ DFT framework must be supplemented with extra consistency conditions. However linearity in the e.o.m of $g_{\mu \nu}$ cannot be achieved when supersymmetry is turned on. 

The present results open the door to future directions:
\begin{itemize}

\item \textbf{$\alpha'$ Corrections}

Finding all the $2$-derivative deformations to the DFT action was addressed in \cite{Tduality} and then fully studied in several works \cite{Correc}. In \cite{Tduality}, a biparametric family of duality covariant theories was introduced. Some of them are low energy effective field theories of string theories but some of them are not (the main example is the so-called HSZ theory \cite{HSZ}). Exploring the Generalized Kerr-Schild ansatz in all these theories is straightforward with the results of this work.   

\item \textbf{Extended Kerr-Schild}

Extended Kerr-Schild (xKS) \cite{xKS} is a possible deformation of the Kerr-Schild anzast which consists in a linear perturbation using $2$ null vectors and the inverse metric tensor receives an exact and second-order perturbation. Implementing this kind of more general but exact ansatz in the context of ${\cal N}=1$ DFT would allow to describe a wide range of heterotic supergravity solutions in a duality covariant way.   

\item \textbf{Classical Double Copy} 

The conventional Kerr-Schild formalism is used to extend the double copy structure of gravity scattering amplitudes to the level of the classical equations of motion \cite{Dcopy}. In \cite{KL} the massless sector of supergravity is included based on the generalized Kerr-Schild ansatz and some aspects of the compatibility of the classical double copy and supersymmetric flat backgrounds were discussed. The present work introduces a way to explore the relation between the classical double copy and general supersymmetric backgrounds associated with the field content of heterotic supergravity.

\item \textbf{Maximal supersymmetry} 

The proper framework to address the generalized Kerr-Schild ansatz in duality covariant theories with maximal supersymmetry is Exceptional Field Theory (EFT)\cite{EFT}. In this context it would be possible to consider a generalized and maximal supersymmetric Kerr-Schild ansatz in a $d=3$ space-time with $E8(8)$ duality, a $d=4$ space-time with E7(7) duality or a $d=5$ space-time with $E6(6)$
symmetry. The way to uplift the formalism considered here to the maximal theory is not straightforward since the field content of these kind of theories are not multiplets of the exceptional groups and compatibility between dualization and the generalized Kerr-Schild ansatz must be firstly studied.  

\end{itemize}
\subsection*{Acknowledgements}
We sincerely thank C.N\'unez, D.Marques,  S.Iguri and T.Codina for many useful comments. We are in debt to K. Lee for exceedingly interesting remarks and comments after the first version of the work. Support by CONICET is also gratefully acknowledged.

\appendix
\section{Background equations of motion}
\label{appendix}

Here we present the equations of motion for the background field content of heterotic supergravity obtained from generic variations of \eqref{BdRAction}. We start with the bosonic sector.
\bea
\Delta\phi_{o} & = & -2\mathcal{L} + 8\ov{\psi}_{o\mu}\gamma^{\nu}\gamma^{\mu}\lambda_{o}D_{\nu}\phi_{o} - 4D_{\nu}\ov{\psi}_{o\mu}\gamma^{\nu}\gamma^{\mu}\lambda_{o} - 4\ov{\psi}_{o\mu}\gamma^{\nu}\gamma^{\mu}D_{\nu}\lambda_{o} - 4\ov{\psi}_{o \mu}\gamma^{\mu}\psi_{o}^{\nu}D_{\nu}\phi_{o}\, \nn \\ 
& & + 2D_{\nu}\ov{\psi}_{o\mu}\gamma^{\mu}\psi_{o}^{\nu} + 2\ov{\psi}_{o\mu}\gamma^{\mu}D_{\nu}\psi_{o}^{\nu}\, \\
\Delta G_{o\mu\nu} & = & \frac{1}{4}g_{o\mu\nu}\Delta\phi_{o} + R_{o\mu\nu} + 4D_{\mu}\phi_{o}D_{o\nu}\phi_{o} - \frac{1}{4}H_{o\mu\lambda\rho}H_{o\nu}{}^{\lambda\rho} - \frac{1}{2}F_{o\mu\lambda i}F_{o\nu}{}^{\lambda i}\, \nn \\
& & - 2g_{o\mu\nu}\ov{\psi}_{o\rho}\gamma^{\lambda}\gamma^{\rho}\lambda_{o}D_{\lambda}\phi_{o} + g_{o\mu\nu}D_{\lambda}\ov{\psi}_{o\rho}\gamma^{\lambda}\gamma^{\rho}\lambda_{o} + g_{o\mu\nu}\ov{\psi}_{o\rho}\gamma^{\lambda}\gamma^{\rho}D_{\lambda}\lambda_{o} + g_{o\mu\nu}\ov{\psi}_{o\rho}\gamma^{\rho}\psi_{o}^{\lambda}D_{\lambda}\phi_{o}\, \nn \\ 
& & - \frac{1}{2}g_{o\mu\nu}D_{\lambda}\ov{\psi}_{o\rho}\gamma^{\rho}\psi_{o}^{\lambda} - \frac{1}{2}g_{o\mu\nu}\ov{\psi}_{o\rho}\gamma^{\rho}D_{\lambda}\psi_{o}^{\lambda} - \ov{\psi}_{o(\mu}\gamma_{\nu)}{}^{\lambda\rho}D_{\lambda}\psi_{o\rho} + \ov{\psi}_{o\lambda}\gamma_{(\mu}{}^{\lambda\rho}D_{\nu)}\psi_{o\rho}\, \nn \\
& & - \ov{\psi}_{o\lambda}\gamma_{(\mu|}{}^{\lambda\rho}D_{\rho}\psi_{o|\nu)} + 4\ov{\lambda}_{o}\gamma_{(\mu}{}^{\rho}D_{\nu)}\psi_{o\rho} - 4\ov{\lambda}_{o}\gamma_{(\mu|}{}^{\rho}D_{\rho}\psi_{o|\nu)} + 4\ov{\lambda}_{o}\gamma_{(\mu}D_{\nu)}\lambda_{o} - \frac{1}{2}\ov{\chi}_{o}^{i}\gamma_{(\mu}D_{\nu)}\chi_{o i}\, \nn \\
& & + 4\ov{\psi}_{(\mu}\gamma^{\rho}\gamma_{\nu)}\lambda_{o}D_{\rho}\phi_{o} + 4\ov{\psi}_{o\rho}\gamma_{(\mu|}\gamma^{\rho}\lambda_{o}D_{|\nu)}\phi - 2\ov{\psi}_{(\mu}\gamma_{\nu)}\psi_{o}^{\rho}D_{\rho}\phi_{o} - 2\ov{\psi}_{o\rho}\gamma^{\rho}\psi_{o(\mu}D_{\nu)}\phi_{o}\, \nn \\ 
& & - \frac{1}{4}\ov{\chi}_{oi}\gamma_{(\mu|}\gamma^{\lambda\rho}F_{o\lambda\rho}{}^{i}\left(\psi_{o|\nu)} + \frac{1}{3}\gamma_{|\nu)}\lambda_{o}\right) - \frac{1}{2}\ov{\chi}_{oi}\gamma^{\lambda}\gamma_{(\mu|}{}^{\rho}F_{o|\nu)\rho}{}^{i}\left(\psi_{o \lambda} + \frac{1}{3}\gamma_{\lambda}\lambda_{o}\right)\, \nn \\
& & + \frac{1}{8}H_{o(\mu|}{}^{\sigma\tau}\left(\ov{\psi}_{o\rho}\gamma^{[\rho}\gamma_{|\nu)\sigma\tau}\gamma^{\lambda]}\psi_{o\lambda} + 4\ov{\psi}_{o\rho}\gamma^{\rho}{}_{|\nu)\sigma\tau}\lambda_{o} - 4\ov{\lambda}_{o}\gamma_{|\nu)\sigma\tau}\lambda_{o} + \frac{1}{2}\ov{\chi}_{o}^{i}\gamma_{|\nu)\sigma\tau}\chi_{oi}\right)\, \nn \\
& & + \frac{1}{24}H_{o\rho\sigma\tau}\left(\ov{\psi}_{o\mu}\gamma_{[\nu}\gamma^{\rho\sigma\tau}\gamma_{\lambda]}\psi_{o}^{\lambda} - \ov{\psi}_{o}^{\lambda}\gamma_{[\nu}\gamma^{\rho\sigma\tau}\gamma_{\lambda]}\psi_{o\mu} + 4\ov{\psi}_{o\mu}\gamma_{\nu}{}^{\rho\sigma\tau}\lambda_{o}\right)\, \\
\Delta b_{o\mu\nu} & = & - D^{\rho}\phi_{o}H_{o\rho\mu\nu} + \frac{1}{2}D^{\rho}H_{o\rho\mu\nu}\, \nn \\ 
& & + \frac{1}{4}D^{\rho}\phi_{o}\left(\ov{\psi}_{o\lambda}\gamma^{[\lambda}\gamma_{\rho\mu\nu}\gamma^{\sigma]}\psi_{o\sigma} + 4\ov{\psi}_{o}^{\lambda}\gamma_{\lambda\rho\mu\nu}\lambda_{o} - 4\ov{\lambda}_{o}\gamma_{\rho\mu\nu}\lambda_{o} + \frac{1}{2}\ov{\chi}_{o}^{i}\gamma_{\rho\mu\nu}\chi_{oi}\right)\, \nn \\
& & - \frac{1}{8}D^{\rho}\left(\ov{\psi}_{o\lambda}\gamma^{[\lambda}\gamma_{\rho\mu\nu}\gamma^{\sigma]}\psi_{o\sigma} + 4\ov{\psi}_{o}^{\lambda}\gamma_{\lambda\rho\mu\nu}\lambda_{o} - 4\ov{\lambda}_{o}\gamma_{\rho\mu\nu}\lambda_{o} + \frac{1}{2}\ov{\chi}_{o}^{i}\gamma_{\rho\mu\nu}\chi_{oi}\right)\, \\
\Delta A_{o\mu}{}^{i} & = & A_{o\rho}{}^{i}\Delta b_{o}^{\rho}{}_{\mu} + \frac{1}{2}H_{o\mu}{}^{\nu\rho}F^{i}_{o\nu\rho} - 2D^{\nu}\phi_{o} F^{i}_{o\nu\mu} + D^{\nu}F^{i}_{o\nu\mu} - \frac{1}{2}\ov{\chi}_{o}^{j}\gamma^{\mu}\chi_{o}^{k}f^{i}{}_{jk}\, \nn \\ 
& & - D^{\nu}\phi_{o}\left(\ov{\chi}_{o}^{i}\gamma_{\rho}\gamma_{\nu\mu}\left(\psi_{o}^{\rho} + \frac{1}{3}\gamma^{\rho}\lambda_{o}\right)\right) + \frac{1}{2}D^{\nu}\left(\ov{\chi}_{o}^{i}\gamma_{\rho}\gamma_{\nu\mu}\left(\psi_{o}^{\rho} + \frac{1}{3}\gamma^{\rho}\lambda_{o}\right)\right)\, \nn \\
& & - \frac{1}{8}F_{o}^{\nu\rho i}\left(\ov{\psi}_{o\sigma}\gamma^{[\sigma}\gamma_{\mu\nu\rho}\gamma^{\lambda]}\psi_{o\lambda} + 4\ov{\psi}_{o}^{\sigma}\gamma_{\sigma\mu\nu\rho}\lambda_{o} - 4\ov{\lambda}_{o}\gamma_{\mu\nu\rho}\lambda_{o} + \frac{1}{2}\ov{\chi}_{o}^{j}\gamma_{\mu\nu\rho}\chi_{oj}\right)
\eea

The equations of motion of the (adjoint) fermionic fields are
\bea
\Delta\ov{\psi}{}_{o}^{\mu} & = & - \gamma^{\mu\nu\rho}D_{\nu}\psi_{o\rho} - 2\gamma^{\mu\nu\rho}\psi_{o\nu}D_{\rho}\phi_{o} + \gamma^{\mu\nu\rho}D_{\rho}\psi_{o\nu} + 4\gamma^{\mu}\gamma^{\nu}\lambda_{o}D_{\nu}\phi_{o}\, \nn \\ 
& & - 4\gamma^{\mu\nu}D_{\nu}\lambda_{o} - 2\gamma^{\mu}\psi_{o}^{\nu}D_{\nu}\phi_{o} + 2\gamma^{\nu}\psi_{o\nu}D^{\mu}\phi_{o} - \frac{1}{4}\gamma^{\mu}\gamma^{\nu\rho}\chi_{oi}F_{o\nu\rho}^{i}\, \nn \\
& & - \gamma^{\nu}\chi_{oi}F_{o\nu}{}^{\mu i} - \frac{1}{16}H_{o}^{\mu\rho\sigma}\left(\gamma_{\nu\rho\sigma}\psi_{o}^{\nu} + 8\gamma_{\sigma}\psi_{o\rho}\right)\, \nn \\
& & - \frac{1}{48}H_{o\rho\sigma\tau}\left(4\gamma^{\nu\rho\sigma\tau\mu}\psi_{o\nu} - 3\gamma^{\sigma\tau\mu}\psi_{o}^{\rho} - 8\gamma^{\mu\rho\sigma\tau}\lambda_{o}\right)\, \\
\Delta\lambda_{o} & = & 4\gamma^{\mu\nu}D_{\mu}\psi_{o\nu} + 8\gamma^{\mu}D_{\mu}\lambda_{o} - 8\gamma^{\mu}\lambda_{o}D_{\mu}\phi_{o} + 4\gamma^{\mu}\gamma^{\nu}\psi_{o\mu}D_{\nu}\phi_{o}\, \nn \\
& & - \frac{1}{12}F_{o\nu\rho}{}^{i}\ov{\chi}_{oi}\gamma^{\mu}\gamma^{\nu\rho}\gamma_{\mu} + \frac{1}{6}H_{o\rho\sigma\tau}\left(\gamma^{\mu\rho\sigma\tau}\psi_{o\mu} - 2\gamma^{\rho\sigma\tau}\lambda_{o}\right)\, \\
\Delta\chi_{o}^{i} & = & \gamma^{\mu}\chi_{o}^{i}D_{\mu}\phi_{o} - \gamma^{\mu}D_{\mu}\chi_{o}^{i} - \frac{1}{4}\gamma^{\mu\nu\rho}\psi_{o\mu}F_{o\nu\rho}^{i} - \frac{1}{2}\gamma^{\rho}\psi_{o}^{\nu}F_{o\nu\rho}^{i}\, \nn \\ 
& & - \frac{1}{3}\gamma^{\nu\rho}\lambda_{o}F_{o\nu\rho}^{i} + \frac{1}{24}H_{o\rho\sigma\tau}\gamma^{\rho\sigma\tau}\chi_{o}^{i}
\eea

\end{document}